\documentclass[aps,prl,onecolumn,longbibliography]{revtex4-1}

\usepackage{mathtools}
\usepackage{amsmath}
\usepackage{amssymb}
\usepackage[margin=1in]{geometry}
\usepackage{color}
\usepackage{graphicx}
\usepackage{natbib}
\usepackage{mathrsfs}
\usepackage{mathtools}
\usepackage{lipsum}
\usepackage[columnwise]{lineno}
\usepackage{comment}
\usepackage[makeroom]{cancel}

\usepackage[normalem]{ulem}

\usepackage[dvipsnames]{xcolor}
\usepackage[colorlinks=true,
            linkcolor=blue,
            urlcolor=blue,
            citecolor=black]{hyperref}

\begin{document}

\newcommand{\ovl}[1]{\overline{#1}}
\newcommand{\br}[1]{\left( #1 \right)}
\newcommand{\DR}{\Delta R_0}
\newcommand{\DP}{\Delta P_0}
\newcommand{\DA}{\Delta A}
\newcommand{\DB}{\Delta B}
\newcommand{\com}[1]{{\color{cyan}[#1]}}
\newcommand\underrel[2]{\mathrel{\mathop{#2}\limits_{#1}}}


\title{The diverse nature of small-scale turbulence}

\author{Ke-Qi Ding,$^{1,2}$ Kun Yang,$^{3}$ Xiang I A Yang,$^4$\footnote[1]{Correspondence: xzy48@psu.edu} Yi-Peng Shi,$^{3,5}$\footnote[2]{Correspondence: ypshi@coe.pku.edu.cn}, and Shi-Yi Chen$^{1,2,5}$\\
\vspace{2mm}
{\small 
$^1$Department of Mechanics and Aerospace Engineering, Southern University of Science and Technology, Shenzhen, Guangdong, 518055, People's Republic of China\\
$^2$Southern Marine Science and Engineering Guangdong Laboratory, Guangzhou, 511458, People's Republic of China\\
$^3$Academy for Advanced Interdisciplinary Studies, Southern University of Science and Technology, Shenzhen, Guangdong 518055, People's Republic of China\\
$^4$ Mechanical Engineering, Pennsylvania State University, State College, Pennsylvania, 16803, USA\\
$^5$ State Key Laboratory for Turbulence and Complex Systems, Peking University, Beijing 100871, People's Republic of China\\}
}


\begin{abstract}
The self-similar Richardson cascade admits two logically possible scenarios of small-scale turbulence at high Reynolds numbers.
In the first scenario, eddies' population densities vary as a function of eddies' scales.
As a result, one or a few eddy types dominate at small scales, and small-scale turbulence lacks diversity.
In the second scenario, eddies' population densities are scale-invariant across the inertial range, resulting in small-scale diversity.
{That is, there are as many types of eddies at the small scales as at the large scales.}
In this letter, we measure eddies' population densities in three-dimensional isotropic turbulence and determine the nature of small-scale turbulence.
The result shows that eddies' population densities are scale-invariant.
\end{abstract}



\maketitle


While turbulent flows are different from one another at large scales, they are universal at small scales. 
Understanding the nature of small-scale turbulence is at the center of turbulence research \cite{frisch1995turbulence, sreenivasan1997phenomenology,johnson2018predicting}.
The beginning point is usually the Richardson cascade \cite{richardson2007weather, kolmogorov1941local}, according to which large-scale eddies break into small-scale eddies, and small-scale eddies break into lesser-scale eddies.
This eddy breakup process is self-similar in the inertial range, where neither viscosity nor flow geometry plays an important role in determining flow's dynamics.
While most authors acknowledge the eddy breakup process as being self-similar, how one models the eddy breakup process differs, and that has led to vastly different speculations about the nature of small-scale turbulence.
Kolmogorov \cite{kolmogorov1941local} models eddy breakup as an even partition of mother eddy's turbulent kinetic energy.
It follows from Kolmogorov that eddies' population densities are scale-invariant, and relatively small-scale turbulence is no different from relatively large-scale turbulence.
On the other hand, Frisch \cite{frisch1978simple} argues that turbulence occupies less space as the cascade process continues to small scales, and small-scale turbulence consists of bursts of velocity fluctuations.
According to Frisch, eddies' population densities are scale-dependent, and the probability of observing turbulence diminishes at small scales.
{This picture was adopted in the study of vortex filaments: vortex filaments occupy less physical space at smaller scales \cite{jimenez1993structure,jimenez1998characteristics}.}
Besides Kolmogorov and Frisch, many have proposed models for the Richardson cascade \cite{benzi1984multifractal, meneveau1987simple, benzi1991multifractality, sreenivasan1991fractals, biferale2004multifractal}.
Like Kolmogorov and Frisch, while they all invoke the Richardson cascade, their models lead to different speculations about the nature of small-scale turbulence.

The Richardson cascade being self-similar says very little about eddies population densities and the nature of small-scale turbulence.
The self-similar Richardson cascade requires the eddy population density scales as {$P(S_i(l))\sim l^{\zeta_i}$} but with no further requirement on $\zeta_i$'s values. 
Here, $S_i(l)$ is a given type of $l$-scaled eddy, $i$ indexes all types of eddies, $P(S_i(l))$ is the probability density function for observing $S_i(l)$, and $\zeta_i$ is a positive number.
In fact, for any $\zeta_i$, $P(S_i(l))$'s variation from one scale $l$ to the next scale $l/2$ is
\begin{equation}
    1-\frac{P(S_i(l/2))}{P(S_i(l))}=1-\frac{1}{2^{\zeta_i}},
\end{equation}
i.e., not a function of $l$ and therefore self-similar irrespective of $\zeta_i$'s value.
{Here, the length scale $l$ in the scaling $P(S_i)\sim l^{\zeta_i}$ needs normalization.
Following the convention, if a process leads to a scaling that is an increasing function of $l$, i.e., if $\zeta>0$, the proper normalization length scale should be the Kolmogorov length scale $\eta$.
The resulting scaling would be $(l/\eta)^{\zeta_i}$.
Consequently, the integral length scale would not be a part of the scaling.
Here, $\zeta_i\geq 0$, and therefore $P(S_i)\sim (l/\eta)^{\zeta_i}$.
In the following, we will omit $\eta$ for brevity.
}
Unlike the Richardson cascade and its insensitivity to $\zeta_i$'s value, the small-scale turbulence and its nature critically depend on whether $\zeta_i$'s are zero.
Consider two eddy types: $i$ and $j$. 
If $\zeta_i\neq \zeta_j$, the fact that $\lim_{l/\eta\to \infty, Re\to \infty}l^{\zeta_i}/l^{\zeta_j}$ is either 0 or infinity suggests that one eddy type dominates the other at small scales.
Here, $\eta$ is the Kolmogorov length scale, and $Re$ is the {Taylor microscale Reynolds number}.
Hence, if $\zeta_i\not \equiv 0$, one or a few eddy types dominate at small scales.
On the other hand, if $\zeta_i\equiv 0$, eddies' population densities are invariant across the inertial range, and there would be as many types of eddies at small scales as at large scales.

Eddies' population density being scale-invariant in the inertial range is, to date, unconfirmed speculation about small-scale turbulence.
It is also a fundamental property of fractal interpolation \cite{scotti1997fractal,scotti1999fractal,basu2004synthetic,ding2010synthetic} (and an implied property of turbulence in Refs \cite{de2013multiscale,wu2020high}).
When applying fractal interpolation, one re-scales the large-scale flows and populates them at small scales, which results in scale-invariant eddy population densities.
However, those fractal models lack {\it a priori} validation, and the question remains open as to what is the true nature of small-scale turbulence.

To answer the above question, we need to measure eddies' population densities, $P(S_i(l))$.
Directly measuring eddies' population densities $P(S_i(l))$ as a function of $l$ is very difficult, if not impossible, as there are many eddy types.
In this letter, we infer $P(S_i(l))$'s $l$ scaling by studying the statistics properties of ``equivalent eddies classes''.
After five steps' derivation, we will come to the conclusion $P(S_i(l))\sim l^0$.


First, we define eddies and eddy classes.
We begin by defining an observation window.
Denote a point in the turbulent flow field as ${\bf x}=(x_1,x_2,x_3)$.
We define $\Omega(l,{\bf x})$ to be a one-dimensional observation window in an arbitrary direction (note that small-scale turbulence is isotropic).
The size of the observation window is $l$, and the point {\bf x} belongs to $\Omega$.
How the size of the observation window is measured can be somewhat arbitrary.
For this discussion, we may think of the observation window as a lens centered at ${\bf x}$ with its length being $l$.
We define a turbulent eddy as the velocity segment within an observation window, i.e., $\{\left.{\bf u}({\bf x})\right|x\in \Omega(l,{\bf x})\}$.
{Thus defined eddies exist everywhere in the flow (as opposed to vortex filaments, which occupy a fraction of the physical space).}
The ensemble of velocity segments at all locations and all scales (all $l$) contains all eddies.
{The definition concerns eddies in the spatial domain only.
We can also define eddies in the temporal domain, and if Taylor's hypothesis holds, we should come to the same conclusions.
}

\begin{figure}
\includegraphics[width=0.49\textwidth]{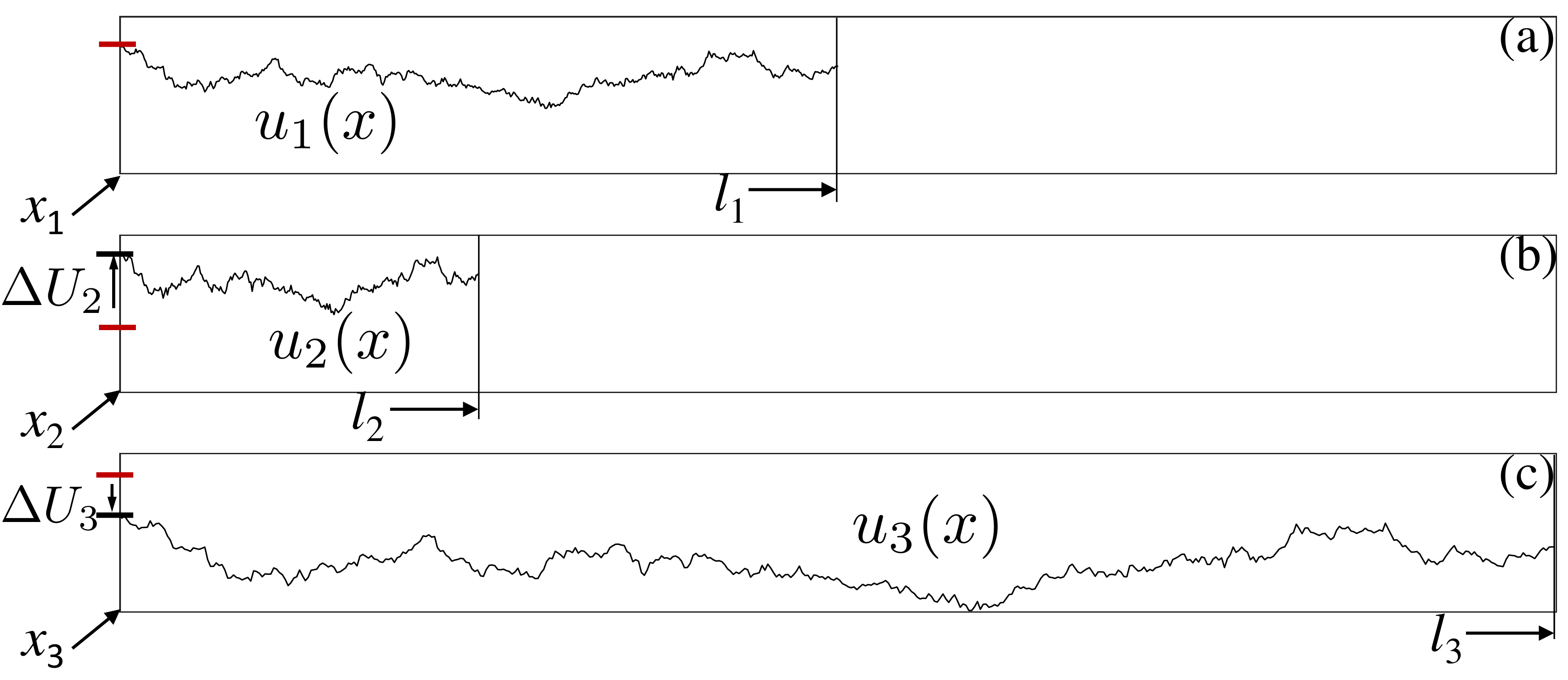}
\caption{Three velocity segments that belong to the same equivalent eddy class.
(b) is (a) compressed and displaced, i.e., $u_2(x_2+l_2(x'-x_1)/l_1)=u_1(x')+\Delta U_2$.
(c) is (a) stretched and displaced, i.e., $u_3(x_3+l_3(x'-x_1)/l_1)=u_1(x')+\Delta U_3$.
} 
\label{fig:eec}
\end{figure}

Second, we define geometric equivalence.
Consider two velocity segments $\{{\bf u}({\bf x'}_1),{\bf x'}_1$ $\in \Omega(l_1,{\bf x}_1)\}$ and $\{{\bf u}({\bf x'}_2),{\bf x'}_2\in\Omega(l_2,{\bf x}_2)\}$. 
We say that the two velocity segments are equivalent if there exist a constant velocity vector ${\bf u_0}$ and a constant positive real number $c$ such that for all $x_1'\in \Omega(l_1,x_1)$, we have
\begin{equation}
{\bf u}({\bf x}_2+l_2({\bf x}'_1-{\bf x}_1)/l_1)=c{\bf u}({\bf x'}_1)+{\bf u_0}.
\label{sim}
\end{equation}
The notion of equivalence makes it possible for us to split velocity segments into equivalent eddy classes \cite{devlin2003sets}, and we denote these equivalent eddy classes as $S_i$, $i=1$, 2, 3, ...
Per our definition, we have:
first, any velocity segment must belong to some equivalent eddy class;
second, two velocity segments that are equivalent must belong to the same equivalent eddy class;
third, one velocity segment cannot belong to two equivalent eddy classes, i.e., equivalent eddy classes are mutually exclusive.
It therefore follows that 
\begin{equation}
    P\left(\cup_{i\in I} S_i(l)\right)= \sum_{i\in I}P\left(S_i(l)\right),
    \label{eq:P}
\end{equation}
for any union of eddy classes $I$.
{Note that the above definition does not concern eddies' dynamics \cite{johnson2016large,johnson2020energy}, and we do not study interactions among eddy classes.}
Figure \ref{fig:eec} shows a few velocity segments that belong to the same equivalent eddy class.

\begin{figure}
\includegraphics[width=0.49\textwidth]{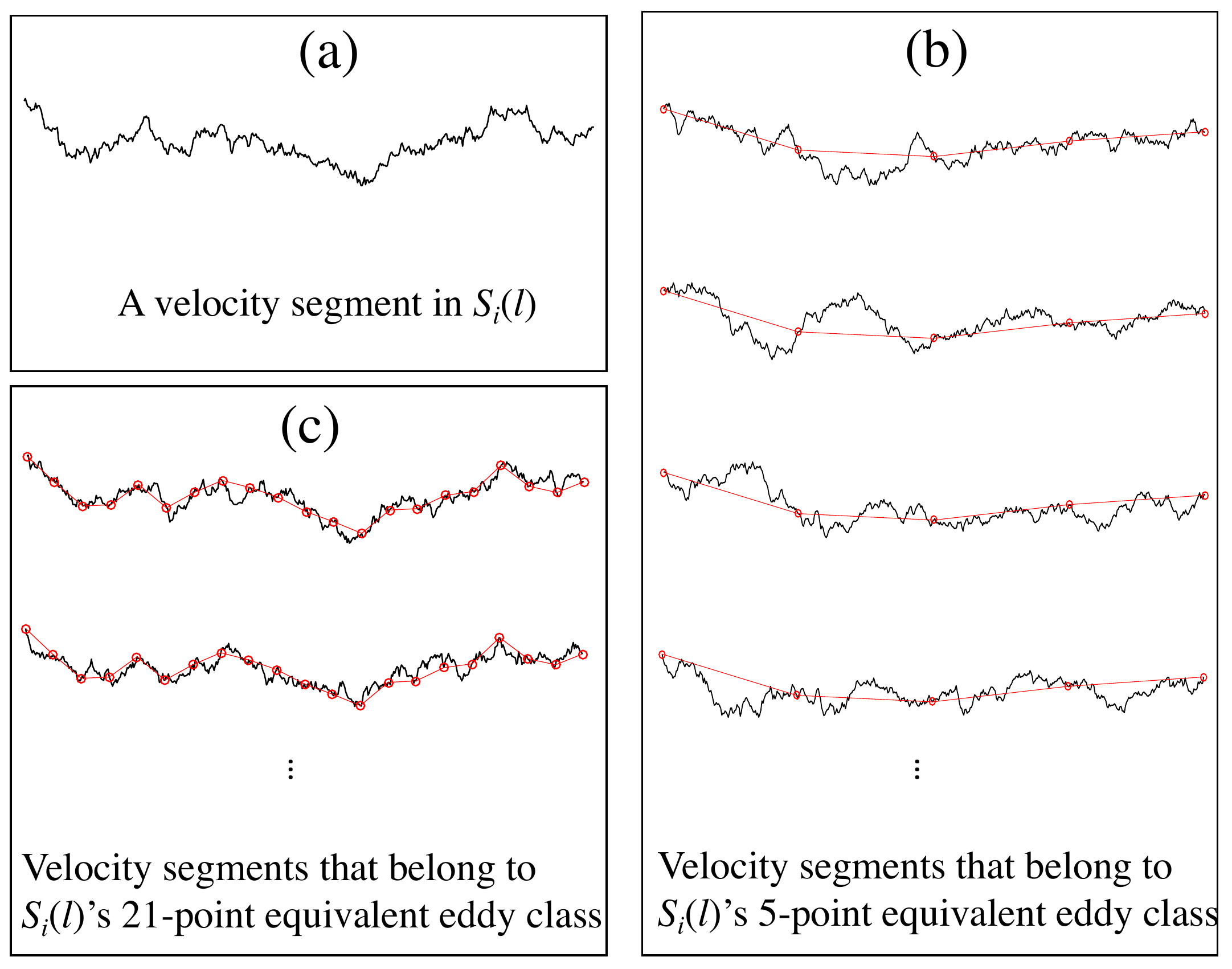}
\caption{\label{fig:npt} (a) A velocity segment in a given $S_i(l)$ and (b, c) $S_i^{(n)}$ for $n=$ 5, and 21.
} 

\end{figure}

Third, we define  $S_i(l)$'s ``$n$-point equivalent eddy class'': {\small $S_i^{(n)}(l)$}.
Given a velocity segment that belongs to $S_i(l)$ and $n$ sampling points on the segment, $S_i(l)$'s $n$-point equivalent eddy class contains all velocity segments that match the given velocity segment at these $n$ sampling points (up to a constant displacement and a multiplying factor).
For example, given the velocity segment  in figure \ref{fig:npt} (a) and its equivalent eddy class $S_i$, the velocity segments in figure \ref{fig:npt} (b, c) belong to $S_i(l)$'s 5-point, and 21-point equivalent eddy classes.
Considering that two sampling points are practically one if the distance between them is less than one Kolmogorov length scale, we require that any two of the $n$ points have a distance of at least one Kolmogorov length scale.
{Considering that the flow is isotropic}, we can think of the $n$ sample points as evenly spaced.
Per the above definition, we have:
first, $S_i(l)$'s $n$-point equivalent eddy class contains $S_i(l)$ itself;
second, as $n$ increases, $S_i(l)$'s $n$-point equivalent eddy class approaches $S_i(l)$ itself;
third, $S_i(l)$ and $S_j(l)$ give rise to the same $n$-point equivalent eddy class if the velocity segments in the two eddy classes $S_i(l)$ and $S_j(l)$ match at the $n$ sampling points;
conversely, if the velocity segments in the two eddy classes $S_i(l)$ and $S_j(l)$ do not match at the $n$-sampling points, their $n$-point equivalent eddy classes, i.e., {\small $S_i^{(n)}(l)$} and {\small$S_j^{(n)}(l)$}, are two different sets;
fourth, the union of all $S_i^{(n)}(l)$ contain all possible velocity segments at the scale $l$.

\begin{figure}
\includegraphics[width=0.49\textwidth]{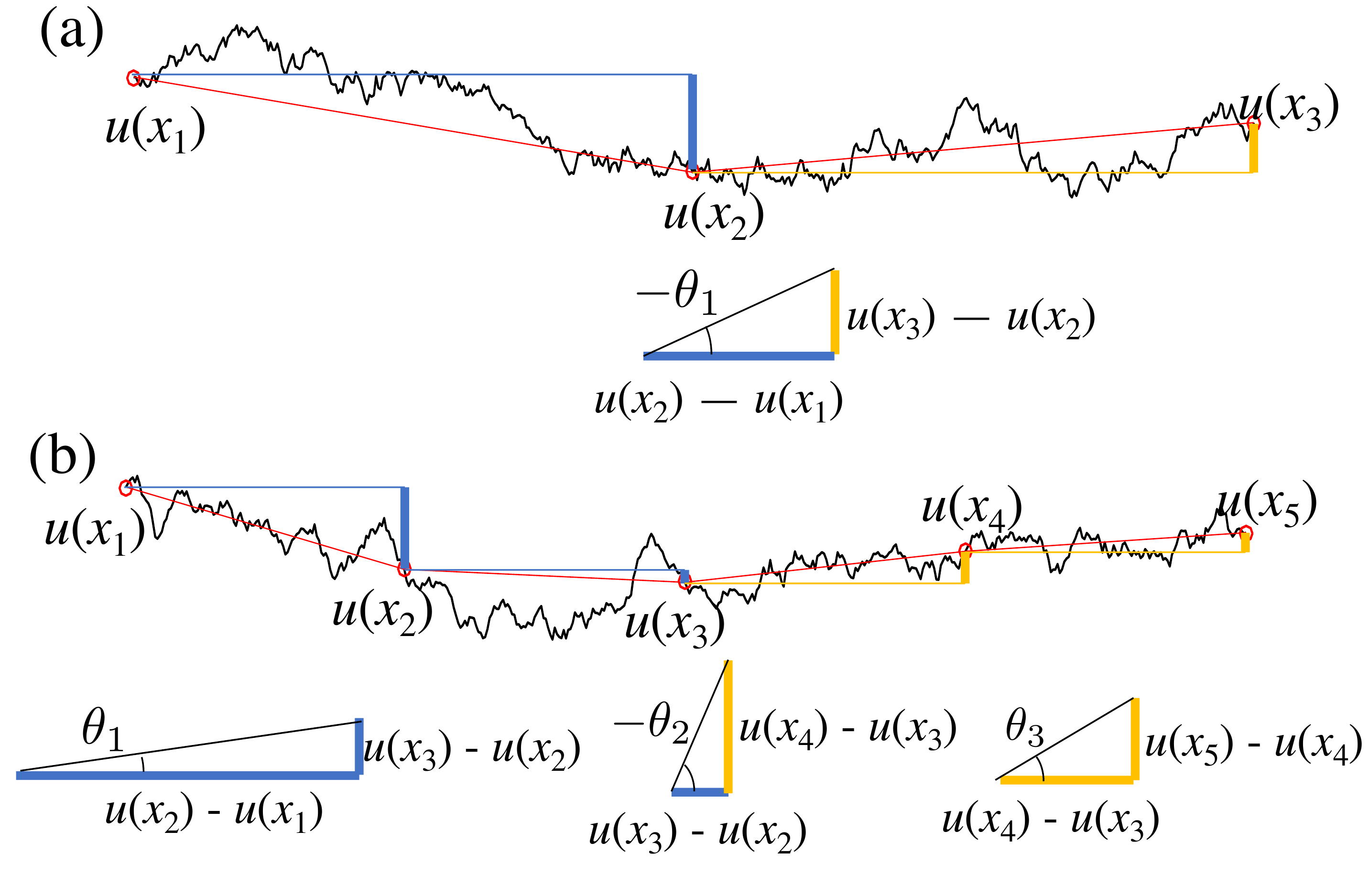}
\caption{\label{fig:theta}Schematic of two velocity segments, and their 3-point and 5-point feature vectors $\pmb{\theta}^{(n)}(l)$.
(a) $\pmb{\theta}^{(3)}(l)$,
(b) $\pmb{\theta}^{(5)}(l)$.
Computing $\theta_i^{(n)}(l)$ involves  $u(x_i)-u(x_{i-1})$ and $u(x_{i-1})-u(x_{i-2})$.
Here, we color $u(x_i)-u(x_{i-1})$ yellow if $u(x_i)>u(x_{i-1})$ and blue if $u(x_i)<u(x_{i-1})$.
Given the definition of $\tan^{-1}$, $\theta_i^{(n)}(l)$ is positive if $\left(u(x_i)-u(x_{i-1})\right)\left(u(x_{i-1})-u(x_{i-2})\right)>0$, and $\theta_i$ is negative if $\left(u(x_i)-u(x_{i-1})\right)\left(u(x_{i-1})-u(x_{i-2})\right)<0$.  
For the two velocity segments here, $\theta_1^{(3)}(l)$ in (a) and $\theta_2^{(5)}(l)$ in (b) are negative; $\theta_1^{(5)}(l)$ and $\theta_3^{(5)}(l)$ in (b) are positive.} 
\end{figure}

In practice, to determine whether a given velocity segment belongs to {\small $S_i^{(n)}(l)$}, we compute the following $n-2$ by 1 feature vector, {\small $\pmb{\theta}^{(n)}(l)$}, whose $i$th component is 
\begin{equation}
    \theta_i^{(n)}(l)=\tan^{-1}\frac{u(x_{i+2})-u(x_{i+1})}{u(x_{i+1})-u(x_i)},
\end{equation}
where $x_{i}$, $i=1$, ..., $n$ is the $i$th sampling point on the velocity segment, $\tan^{-1}$ is the inverse of the tangent function, and we define $\tan^{-1}(\pm\infty)=\pm \pi/2$.
Two velocity segments that give rise to the same $\pmb{\theta}^{(n)}(l)$ belong to the same $n$-point equivalent eddy class $S_i^{(n)}(l)$.
Figure \ref{fig:theta} sketches how one may compute $\pmb{\theta}^{(n)}(l)$ for $n=3$ and $n=5$.

Fourth, we compute $\pmb{\theta}^{(n)}(l)$'s statistics, the knowledge of which will allow us to infer $P(S_i^{(n)}(l))$'s $l$ scaling.
Formally, given a function $f(\pmb{\theta}^{(n)}(l))$, {its ensemble average is its $P(S_i(l))$ weighted sum over all possible $S_i(l)$'s}, and therefore we have
\begin{equation}
\footnotesize
    \left<f\left(\pmb{\theta}^{(n)}(l)\right)\right>
    =\sum_{i'}P\left(S_{i'}^{(n)}(l)\right)\left<f\left(\pmb{\theta}^{(n)}(l)\right)\right>_{S_{i'}^{(n)}(l)},
    \label{eq:ftht}
\end{equation}
where the summation is among the mutually exclusive $n$-point equivalent eddy classes, $\left<\cdot\right>_{S_{i'}^{(n)}(l)}$ is the ensemble average given only velocity segments in $S_{i'}(l)$'s $n$-point equivalent eddy class ${S_{i'}^{(n)}(l)}$.
{While it is not the focus of this work, we can compute any statistics by summing up contributions due to all eddy classes.
For example, the $p$th-order velocity structure function is
\begin{equation}
\begin{split}
    &\left<(u(x+l)-u(x))^p\right>\\
    =&\sum_i P(S_i(l))\left<(u(x+l)-u(x))^p\right>_{S_i(l)}.
\end{split}
\end{equation}}
Here, we compute {\small $f(\pmb{\theta}^{(3)}(l))=(\theta_1^{(3)}(l))^2$} according to Eq. \eqref{eq:ftht}.
Let us say that the flow has only two mutually exclusive 3-point equivalent eddy classes: {\small $S_{1}^{(3)}(l)$} and {\small $S_{2}^{(3)}(l)$}.
The velocity segments in {\small $S_{1}^{(3)}(l)$} and {\small $S_{1}^{(3)}(l)$} correspond to the feature vectors $\theta'$ and $\theta''$.
(For $n=3$, the feature vector has only one component.)
The eddy population densities are {\small $P(S_{1}^{(3)}(l))\sim l^{\zeta_1}$} and {\small $P(S_{2}^{(3)}(l))\sim l^{\zeta_2}$} as required by the Richardson cascade.
It follows from Eq. \eqref{eq:ftht} that
\begin{equation}
\begin{split}
    \left<\left(\theta_1^{(3)}(l)\right)^2\right>
    &\sim c_1(l/\eta)^{\zeta_1}\theta'^2+c_2(l/\eta)^{\zeta_2}\theta''^2,
\end{split}
\label{eq:f3}
\end{equation}
where $c_1$ and $c_2$ are two constants.
If $\zeta_1\neq \zeta_2$, one of the two terms in Eq. \eqref{eq:f3} dominates at sufficiently high Reynolds numbers.
Without loss of generality, let us say $\zeta_1\geq \zeta_2\geq 0$.
For a given $l/L$, we have
\begin{equation}
\begin{split}
\small
    &\lim_{Re\to \infty}\left<\left(\theta_1^{(3)}(l)\right)^2\right>\\
    \sim &\lim_{l/\eta\to \infty} \left({l}/{\eta}\right)^{\zeta_1}\left[1+c_3\left({l}/{\eta}\right)^{\zeta_2-\zeta_1}\right]\\
    = &(l/\eta)^{\zeta_1}.
\end{split}
    \label{eq:f3-1}
\end{equation}
In this case, {\small $\left<(\theta_1^{(3)}(l))^2\right>\sim l^0$} if and only if $\zeta_1=0$.
Also, because $\zeta_1\geq \zeta_2\geq 0$, if $\zeta_1=0$, we would have $\zeta_1=\zeta_2=0$, and the population densities of the two 3-point equivalent eddy classes would have $l^0$ scaling. 
On the other hand, if $\zeta_1=\zeta_2(=\zeta)$, Eq. \eqref{eq:f3} becomes
\begin{equation}
    \left<\left(\theta_1^{(3)}(l)\right)^2\right>
    \sim l^{\zeta}(\theta'^2+\theta''^2).
    \label{eq:f3-2}
\end{equation}
Again, {\small $\left<(\theta_1^{(3)}(l))^2\right>\sim l^0$} if and only if $\zeta=0$.
{The above argument relies on {\it a priori} knowledge of $\zeta_i$'s sign.}
In the supplemental material, we present a derivation that does not rely on our knowledge of $\zeta$'s sign.
{The idea is to consider two $\mathbf{\theta}$'s statistics.
We would then be able to determine $\zeta_{1,2}$'s values directly from the two scalings.
(It is like solving for two unknowns from two equations.)}

Generalizing the above derivation to an arbitrary number of 3-point equivalent eddy classes, Eq. \eqref{eq:f3} becomes
\begin{equation}
    \left<\left(\theta_1^{(3)}(l)\right)^2\right>
    \sim l^{\zeta_1}\theta'^2+l^{\zeta_2}\theta''^2+l^{\zeta_3}\theta'''^2+...
\end{equation}
Following the same logic, we conclude that if {\small $\left<(\theta_1^{(3)}(l))^2\right>\sim l^0$}, the eddy population density scales as {\small $P(S_i^{(3)}(l))\sim l^0$}.

\begin{figure}
\begin{center}
\includegraphics[height=1.7in]{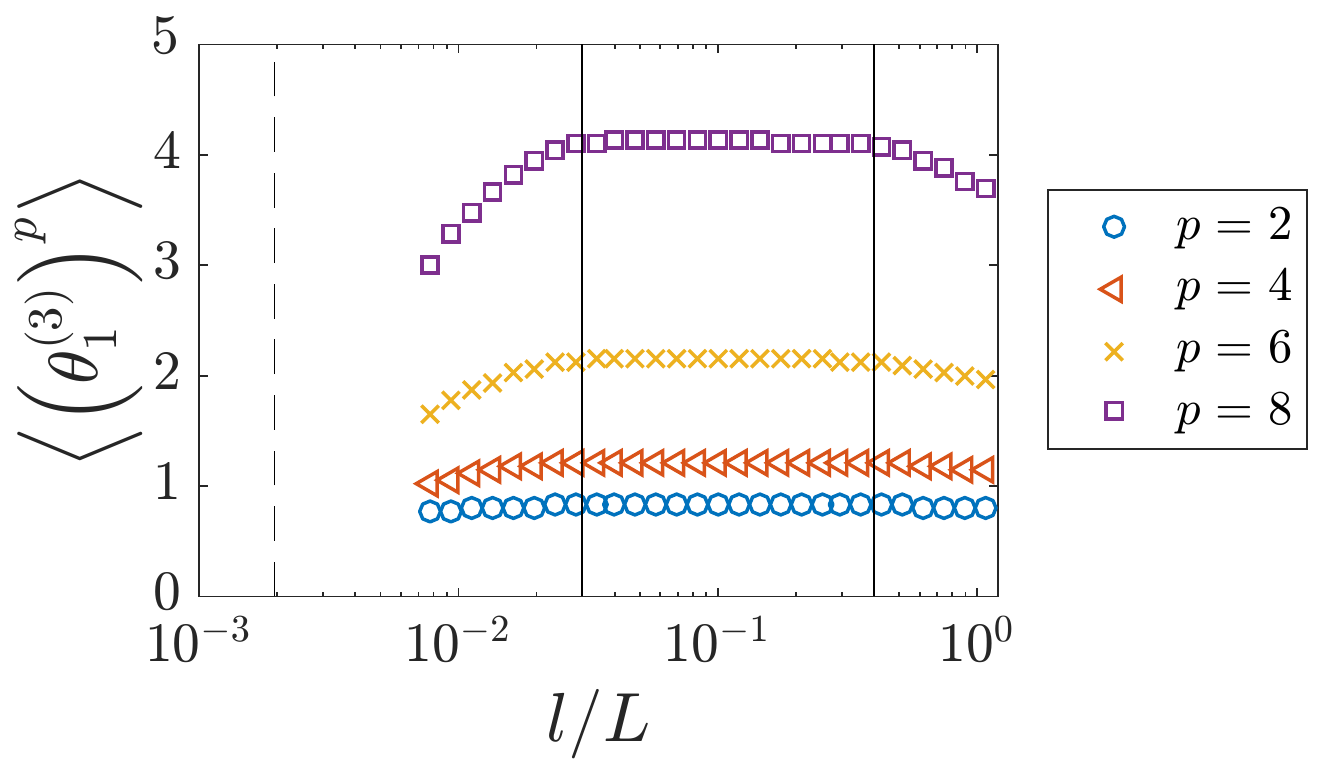}
\caption{ \label{fig:tht1}  $\left<\left(\theta_1^{(3)}(l)\right)^p\right>$ for  $p=2, 4, 6, 8$. 
$L$ is the length of the periodic computational box in one of the three Cartesian directions.
The dashed lines are at the grid cutoff.
The solid lines encompass the inertial range.
} 
\end{center}
\end{figure}

We now examine the data to see if {\small $\left<(\theta_1^{(3)}(l))^2\right>$} scales as $l^0$.
Figure \ref{fig:tht1} shows {\small $\left<(\theta_1^{(3)}(l))^p\right>$} for $p=2, 4, 6, 8$ in a $Re_\lambda=433$ isotropic turbulent flow.
Here, $Re_\lambda$ is the Taylor-scale Reynolds number.
The data is DNS of isotropic turbulence in a periodic box.
The grid size is $1024^3$, and the domain size is $2\pi^3$.
Further details of the DNS data can be found in Ref \cite{cao1999statistics}.
We see that not only {\small $\left<(\theta^{(3)}_1)^2\right>$} scales as $l^0$ in the inertial range but also the higher order even moments.
This allows us to conclude that, for any $i$,
\begin{equation}
\small
    P\left(S_i^{(3)}(l)\right)\sim l^0.
    \label{eq:P3}
\end{equation}

Next, we consider $n$-point equivalent eddy classes $S_i^{(n)}$, whose feature vectors' size is $n-2$ by 1.
We have {\small $\left<(\theta_k^{(n)}(l))^2\right>$}:
\begin{equation}
\small
    \left<\left(\theta_k^{(n)}(l)\right)^2\right>
    \sim c_1l^{\zeta_1}\theta'^2_k+c_2l^{\zeta_2}\theta''^2_k+...,
    \label{eq:fn}
\end{equation}
for $k=1,~2,~3, ...,~n-2$.
Following the same logic, if the data is such  {\small $\left<(\theta_k^{(n)}(l))^2\right>\sim l^0$} for $k=1$, 2, 3, ..., $n-2$, we would be able to conclude $P(S_i^{(n)}(l))\sim l^0$.
To prove {\small $\left<(\theta_k^{(n)}(l))^2\right>\sim l^0$}, we invoke the following two facts:
first, because the flow is homogeneous, for evenly spaced sampling points, we have {\small $\left<\theta_{k'}^{(n)}(l)\right>=\left<\theta_{k''}^{(n)}(l)\right>$} for any $k'$ and $k''$;
second, per our definition, {the segment between the first and the third sampling points of an velocity segment in $S^{(n)}_i(l)$ is a velocity segment in $S^{(3)}_i(2l/(n-1))$,} and therefore 
\begin{equation}
    \left<\left(\theta_1^{(n)}(l)\right)^2\right>\equiv\left<\left(\theta_1^{(3)}\left(\frac{2l}{n-1}\right)\right)^2\right>.
    \label{eq:n-3}
\end{equation}
Hence, to show {\small $\left<(\theta_k^{(n)}(l))^2\right>\sim l^0$} for $k=1$, 2,  3,..., $n-2$, we only need to show {\small $\left<(\theta_1^{(3)}(l))^2\right>\sim l^0$}, which is the result in figure \ref{fig:tht1}.

Fifth (and the last step), we show {\small $P(S_i(l))\sim l^0$}.
This is now trivial. 
Because $S_i^{(n)}(l)$ becomes $S_i(l)$ itself for sufficiently many sampling points, the fact that {\small $P(S_i^{(n)}(l))\sim l^0$} for any $n$ readily guarantees
\begin{equation}
    P\left(S_i(l)\right)\sim l^0,
    \label{eq:Pn}
\end{equation}
and we come to our conclusion.

To summarize, we show that eddies' population density is scale-invariant across the inertial range, i.e., $P(S_i(l))\sim l^0$.
{The result shows that there are as many types of eddies at small scales as at large scales.}






\section{Acknowledgement}
We thank C Meneveau for fruitful discussion. 
Y.-P. Shi is supported by Projects 91752202 from the National Natural Science Foundation of China.

\section{Appendix: a more rigorous derivation}

Let us say that the flow has only two mutually exclusive 3-point equivalent eddy classes: $S_1^{(3)}(l)$ and $S_2^{(3)}(l)$, whose feature vectors are $(\theta')$ and $(\theta'')$ and their eddy population densities scale as $P(S_1^{(3)}(l))\sim l^{\zeta_1}$ and $P(S_2^{(3)}(l))\sim l^{\zeta_2}$.
In order to arrive at the conclusion $\zeta_1=\zeta_2=0$, we assume that $\zeta_{1,2}>0$ in the main text.
In this supplemental material, we present a derivation that does not rely any assumption about $\zeta_{1,2}$'s sign.

\begin{figure}[htb!]
\centering
\includegraphics[width=0.44\textwidth]{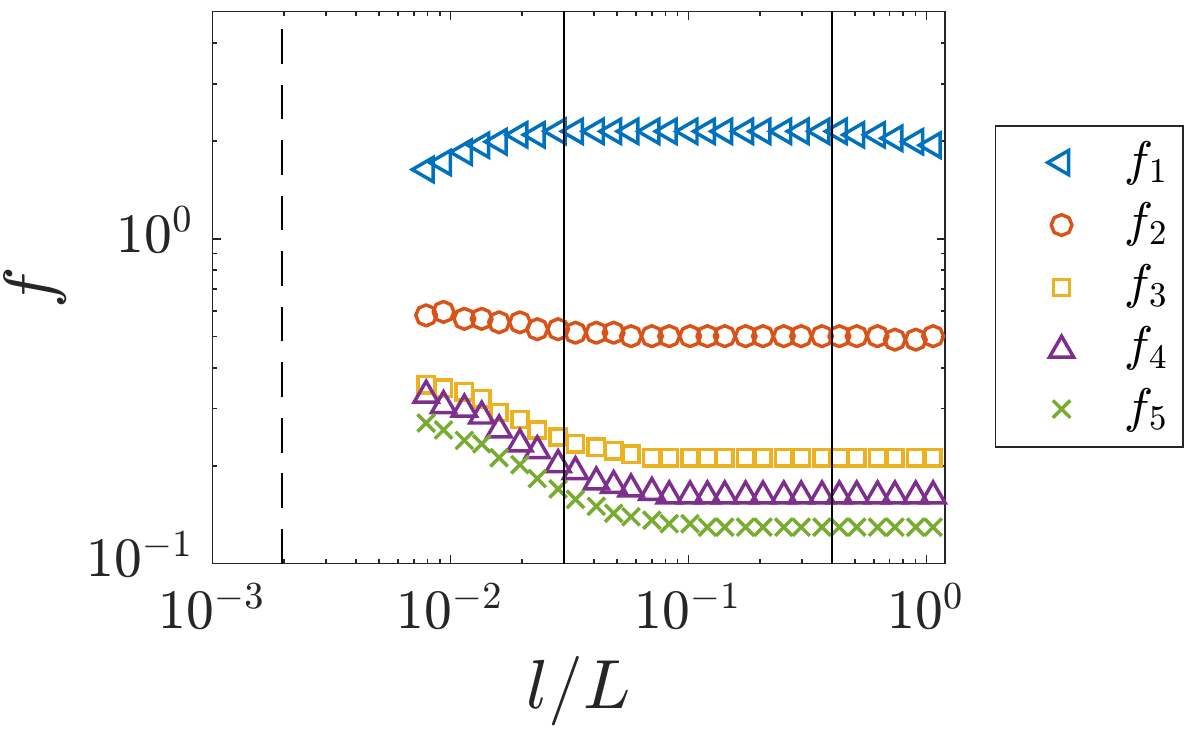}
\caption{Here, $f_1=\left<|\theta_1|^6\right>$,
$f_2=\left<|\theta_1|^3|\theta_2|^3\right>$,
$f_3=\left<|\theta_1|^2|\theta_2|^2|\theta_3|^2\right>$,
$f_4=\left<|\theta_1|^2|\theta_2|^2|\theta_3||\theta_4|\right>$, and
$f_5=\left<|\theta_1|^2|\theta_2||\theta_3||\theta_4||\theta_5|\right>$ for $n=7$ in a $Re_\lambda=344$ isotropic turbulence. 
The dashed line is at the grid cutoff.
The two solid lines enclose the scales within which the energy spectrum follows a $-5/3$ scaling.} 
\label{fig:thtn} 
\end{figure}

We consider two $\pmb{\theta}$ statistics, i.e.,  {\small $\left<(\theta_1^{(3)}(l))^2\right>$} and {\small $\left<(\theta_1^{(3)}(l))^4\right>$} at two arbitrary length scales, $l_1$ and $l_2$:
\begin{equation}
\small
\begin{split}
    \left<(\theta_1^{(3)}(l_1))^2\right>
    = P(S_1^{(3)}(l_1))\theta'^2+P(S_2^{(3)}(l_1))\theta''^2, \\
    \left<(\theta_1^{(3)}(l_1))^4\right>
    = P(S_1^{(3)}(l_1))\theta'^4+P(S_2^{(3)}(l_1))\theta''^4, \\
\end{split}
\label{eq1}
\end{equation}
and
\begin{equation}
\small
\begin{split}
    \left<(\theta_1^{(3)}(l_2))^2\right>
    = P(S_1^{(3)}(l_2))\theta'^2+P(S_2^{(3)}(l_2))\theta''^2, \\
    \left<(\theta_1^{(3)}(l_2))^4\right>
    = P(S_1^{(3)}(l_2))\theta'^4+P(S_2^{(3)}(l_2))\theta''^4, \\
\end{split}
\label{eq2}
\end{equation}
Rewriting Eqs. \eqref{eq1} and \eqref{eq2}, we have
\begin{equation}
    \begin{bmatrix}
    \theta'^2 & \theta''^2\\
    \theta'^4 & \theta''^4
    \end{bmatrix}
    \begin{bmatrix}
    P(S_1^{(3)}(l_1))\\
    P(S_2^{(3)}(l_1))
    \end{bmatrix}
    =
    \begin{bmatrix}
    \left<(\theta_1^{(3)}(l_1))^2\right>\\
    \left<(\theta_1^{(3)}(l_1))^4\right>
    \end{bmatrix}
\label{eq12}
\end{equation}
and 
\begin{equation}
    \begin{bmatrix}
    \theta'^2 & \theta''^2\\
    \theta'^4 & \theta''^4
    \end{bmatrix}
    \begin{bmatrix}
    P(S_1^{(3)}(l_2))\\
    P(S_2^{(3)}(l_2))
    \end{bmatrix}
    =
    \begin{bmatrix}
    \left<(\theta_1^{(3)}(l_2))^2\right>\\
    \left<(\theta_1^{(3)}(l_2))^4\right>
    \end{bmatrix}.
\label{eq22}
\end{equation}
The determinant of the 2-by-2 matrix 
\begin{equation*}
    \begin{bmatrix}
    \theta'^2 & \theta''^2\\
    \theta'^4 & \theta''^4
    \end{bmatrix} 
\end{equation*}
is
\begin{equation}
    \theta'^2\theta''^4-\theta'^4\theta''^2\neq 0
\end{equation}
because $\theta'\neq \theta''$.
Now, if {\small $\left<(\theta_1^{(3)}(l))^2\right>\sim l^0$} and {\small $\left<(\theta_1^{(3)}(l))^4\right>\sim l^0$}, the right hand side of Eqs. \eqref{eq12} and \eqref{eq22} are equal.
As a result,
\begin{equation}
    \begin{bmatrix}
    P(S_1^{(3)}(l_1))\\
    P(S_2^{(3)}(l_1))
    \end{bmatrix}
    =
    \begin{bmatrix}
    P(S_1^{(3)}(l_2))\\
    P(S_2^{(3)}(l_2))
    \end{bmatrix},
\end{equation}
for arbitrary $l_1$ and $l_2$, i.e.,
$P(S_1^{(3)}(l_2))\sim l^0$ and $P(S_2^{(3)}(l_2))\sim l^0$, leading to the conclusion $\zeta_1=\zeta_2=0$.
If we have $n$ mutually exclusive 3-point equivalent eddy classes, we need to show that $\left<(\theta_1^{(3)}(l))^{2n}\right>\sim l^0$, which is shown in figure 4 of the main text.
In fact, data shows that any $\pmb{\theta}$ statistics is scale-invariant within the inertial range.
Figure \ref{fig:thtn} shows a few $\pmb{\theta}^{(n)}(l)$'s statistics.
We see that the statistics scales as $l^0$ in the inertial range. 

\bibliography{reference.bib}

\end{document}